\documentstyle[11pt] {article}

\title{Phase-Modulus Relations for a Reflected Particle}

\author{ A. Yahalom$^a$, R. Englman$^{a,b}$  \\
$^a$ College of Judea and Samaria, Ariel 44284, Israel\\
$^b$ Department of Physics and Applied Mathematics,\\
Soreq NRC,Yavne 81800,Israel\\
e-mail: asya@ycariel.yosh.ac.il; englman@vms.huji.ac.il}

\begin{document}

\maketitle

\newcommand{\beq} {\begin{equation}}
\newcommand{\enq} {\end{equation}}
\newcommand{\ber} {\begin {eqnarray}}
\newcommand{\enr} {\end {eqnarray}}
\newcommand{\eq} {equation}
\newcommand{\eqs} {equations }
\newcommand{\mn}  {{\mu \nu}}
\newcommand{\sn}  {{\sigma \nu}}
\newcommand{\rhm}  {{\rho \mu}}
\newcommand{\sr}  {{\sigma \rho}}
\newcommand{\bh}  {{\bar h}}
\newcommand {\er}[1] {equation (\ref{#1}) }
\newcommand {\SE} {Schr\"{o}dinger equation}
\newcommand {\Del} {\Delta}

\begin {abstract}
We formulate  analytically the reflection of a one dimensional, expanding
free wave-packet (wp) from an infinite barrier. Three types of wp's
 are considered, representing an electron, a molecule and a classical
 object. We derive a threshold criterion for the values of the dynamic
 parameters so that reciprocal (Kramers-Kronig) relations hold
 {\it in the time domain} between the log-modulus of the wp and the (analytic
   part of its) phase acquired during the reflection. For an electron, in
 a typical case, the relations are shown to be satisfied.
 For a molecule the modulus-phase relations take a more
  complicated form, including the so called Blaschke term. For a classical
 particle characterized by a
large mean momentum ($\hbar  K >>  \frac{\hbar ~ trajectory~ length }
{(size~ of~ wave-packet)^2} >>> \frac{\hbar }{size~ of~ wave-packet}$)
the rate of acquisition of the relative phase between different wp components
is enormous (for a bullet it is
typically $10^{14}$ GHertz) with  also a very large value for the phase
maximum.

\end {abstract}
\section {Background and Introduction}
Textbooks of quantum mechanics contain accounts of the impingement of a
freely moving one-dimensional particle on a finite- or an
infinite-height barrier (e.g., ref. 1)   Some further
developments in ref. 2 and more recently in ref. 3
derived the intensity or modulus of the particle beyond a barrier. In
some of these works the particle is modelled by a incoming plane wave with
a given momentum. Other related works  are ref. 4  and, on the
experimental side, ref. 5. In so much as at some later
stage in this study we obtain the location of the zeros of the reflected
wave, we note the recent interest in the distributions of zeros (nodes)
  of a (chaotic) wave~$^{6,7}$ in the coordinate space.
 %In \subsection {dense distribution} we derive the
%distribution of zeros of a reflected wp in the (complex) time-plane.

Using an elementary and exactly soluble model, we consider here a
localized wave packet (wp) representing, for instance, a microscopic
particle, such as an electron and a molecule, reflected from an
infinitely high barrier. One of our purposes is to investigate the phase
of the wp during its history and to unravel a possible relation between
the phase and the modulus of the wp. Such relations were treated in our
earlier articles $~^{8-10}$, and this work can be considered an extension.

 We further consider the wp of a classical particle and look at its
  phase behavior. The
interest in this lies in the widely held  belief  that in any quantum
mechanical measuring-process the phase interference between the measured
quantum system and its classical enclosure plays a crucial role (e.g., ref. 11
). An opposite view has been recently expressed in ref. 12.

\section{Description of the Model}

\begin{figure}
\vspace{8cm}
\begin{picture}(1,1)
\end{picture}
\includegraphics{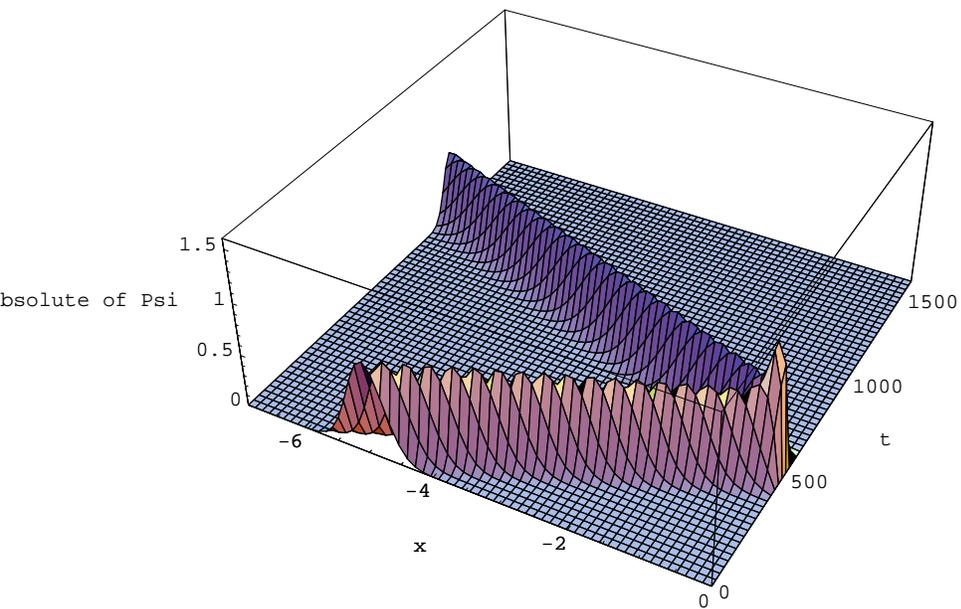}
\caption {Time-motion plot of the wave-packet of a particle reflected by
a boundary placed at $x=0$. The particle starts at $x=-5$. (The units
  employed vary with the nature of the particle: an electron, a molecule
  or a macroscopic projectile.) }
\label{scat}
\end{figure}

The particle is represented as a one-dimensional expanding
wave-packet (wp), starting at time $t=0$ as a Gaussian form  centered at
a point $x=a$ and having initially a width $2\Del$, that is as
\beq
\psi(x,a,t)  = \frac{1}{\sqrt \Del} \exp\big
[-\frac{(x-a)^2}{4\Del^2 }\big]\exp\big[iK(x-a)]
\label{psi0}
\enq
$K$ is defined  in terms of the mean particle momentum $p$ by $K=\frac{p}{\hbar}$
and  the crossed Planck constant $\hbar$.  The particle (whose physical mass
 is $m_{physical}$)  moves to the right, until it
impinges on an infinite barrier at $x=0$ (so that $a$ is negative). It is
then, reflected from the barrier and moves left-wards, as depicted in
figure \ref{scat}.
 Following Tomonaga~$^{13}$
  and other elementary texts, we write the wave-function
$\psi$ of a freely moving wp ({\it without} a barrier) and satisfying a time
 dependent \SE, as
\ber
\psi(x,a,t)&=&\frac{1}{\sqrt (\Del+\frac{it}{2m\Del} )}
exp\big
[-\frac{(x-a)^2-4i\Del ^2K (x-a-\frac{Kt}{2m})}{4\Del^2 +\frac{2it}{m}}\big]\\
          &=&(\frac{1}{\Del^2+\frac {t^2}{4\Del^2m^2}})^{\frac{1}{4}}
 \exp\big [-\frac{(x-a-\frac{Kt}{m})^2}{
4\Del^2+\frac{t^2}{\Del^2 m^2}}\big]
 \exp\big( -\frac{i}{2} arctan \frac{t}
{2\Del^2 m}\big)
\nonumber\\
 & & \exp\big[ i\frac{K(x-a-\frac{Kt}{2m})+\frac{(x-a)^2}
{8 \Del^2}\frac{t}{m\Del^2}}
{1+\frac{t^2}{4\Del^4 m^2}}\big]
\label {psi}
\enr
The symbol $m$ is related to the physical mass by $m=\frac{m_{physical}}{\hbar}$.
In the second expression the wave function appears as a product
of the (real) modulus and a pure phase factor. The group (or amplitude) velocity
is  $\frac{Kt}{m}$, the same as the velocity of a classical particle, but
the phase velocity is seen to be more complicated. It is noted for future reference
that when the above free wp is considered as a function of the complex time
\beq
t=t'+it"
\label {complex}
\enq
it is regular function in the lower half of the complex $t$-plane ($t"<0$)
and tends to zero for
large $|t|$. The former property is shared by all physical wave-packets in a
time-independent environment, and  is due to the
 lower  boundedness of the energy (or frequency) spectrum (which implies that
wave functions of a freely moving particles with negative energies are
all zero $^{14,10}$. This property is
analogous to the principle of causality, which makes the response
functions to be zero at negative times and by consequence, ensures the
analyticity of response functions in the upper half of the complex
frequency plane $^{15}$.

It will be noticed that the wp has a branch point and an essential
singularity in the upper half of
 the $t$- plane.

 In the presence of an infinite barrier the wave function, to be written as
 $\Psi(x,t)$, has to satisfy the  condition
\beq \Psi(x,t)=0~~ \
for ~x=0 \label{bc}
 \enq
 at all times and to vanish
at $x=-\infty$ at finite times. A suitable solution is thus
 \beq
\Psi(x,t)=N [\psi(x,a,t)-\psi(-x,a,t)]~~ for ~x<0
 \label{Psi}
 \enq and
\beq \Psi(x,t)=0 ~~ for ~ x \ge 0
\label{Psimin}
\enq
 The real, positive
quantity $N $ in \er{Psi} is a normalizing factor, given by \beq N
^{-2}=\int_{-\infty}^{0} dx{|\Psi(x,t)|^2}
 \label {normalizer}
  \enq
   and is
independent of time. This result follows from the integration of the
continuity (or mass-conservation) equation, account being taken of the
vanishing of the integrand at $x=-\infty$ and at $x=0$.

\section{The convergent difference function}
The full wave function $\Psi(x,t)$ shown in \er {Psi} can be expressed as
a product of the incoming wave function ($\psi(x,t)$ shown in \er {psi}) and
 the  "difference function" $\chi '(x,t)$ defined as
\beq
\chi '(x,t)=1-\frac{\psi(-x,t)}{\psi(x,t)}
\enq
From \er {psi} this is
\beq
\chi '(x,t) =  1-e ^{\frac{-4x(2iK\Del^2 +a)}{4\Del^2+\frac{2it}{m}}}
\label {chi1}
\enq
This function contains the effect of the reflection on both the amplitude and  the
phase of the total wave function. We now introduce a new function $\chi (x,t)$
, given by
\beq
\chi (x,t)=\frac{4\Del^2+\frac{2it}{m}}{4x(2iK\Del^2 +a)} (1-e
^\frac{-4x(2iK\Del^2 +a)}{4\Del^2+\frac{2it}{m}})
 \label{chi}
 \enq
differing from the (former) difference function $\chi' (x,t)$ only by the fraction
shown.  This
function  has the desired analytical property of tending to 1 as $|t| \to
\infty$. By consequence $\ln \chi (x,t) \to 0$ and we shall be able to use
this function  in an integration of the logarithm over $t$ with infinite
limits in the formulae that follow. $\chi (x,t)$ is thus termed the
"convergent difference function" and has (in  certain physical situations,
 to be specified later) the properties postulated in ref. 8-10
for Hilbert transforms.
 \subsubsection {Reciprocal relations}
 The validity of the
following formulae requires $\ln\chi (x,t)$ to be analytic in the lower
half of the complex t-plane and to tend to zero as $|t| \to \infty $.
\beq \frac{1}{\pi} P \int_{-\infty}^{\infty}dt' [\ln|\chi(t')|]/(t-t') = -
\arg \chi (t) \label {RRim} \enq and \beq \frac{1}{\pi} P
\int_{-\infty}^{\infty}dt' [\arg\chi(t')]/(t-t') = \ln|\chi (t)|
\label{RRre} \enq
 Here P signifies the principal part of the singular integral. (For a
 derivation and extensions of these formulae when not all the conditions
  are met,  see ref. 16 or 9.)

\subsection {Zeros of the difference function}
Evidently $\chi (x,t)$ has no {\it singularities} in the lower half of the complex $t$-plane.
 However, since our interest is in the logarithm, we need to examine not
  only the singularities but also the {\it zeros}
of  $\chi (x,t)$. These will come about when the exponent in \er {Psi} is
zero or an integer times $2\pi $. Writing in the sequel $-|x|$ for $x$
 and $-|a|$ for $a$ (since both quantities are negative) and equating
\beq
\frac{4|x|(2iK\Del^2 -|a|)}{4\Del^2+\frac{2it}{m}} =2i\pi n
\label {zeros}
\enq
(where $n$ is a positive or a negative integer or zero), we find the location
  of the $n$'th zero in the complex $t$-plane
 ($t=t'+it"$), at any fixed location $-|x|$ on the left of the barrier as
 follows:
\ber t_n & = & t'_n + it"_n\\ t'_n & = & m \frac{|x||a|}{n\pi},
~~~t"_n=2\Del^2 m (1-\frac{K|x|}{n\pi}) \label{t_n} \enr
 The time when
the center of the wave is reflected from the barrier is
 denoted by $t_r$ and is
\beq
t_r=\frac{m|a|}{K}
\label{tb}
\enq
whereas the time (to  be denoted by $t_d$) when the wp broadens  due to its
intrinsic dynamics in excess of its original width is given by
\beq
t_d= 2m\Del^2
\label{td}
\enq
We next define for any fixed point $x$ to the left of the barrier the dimensionless quantity
 $n_r(x)$ given by
\beq
n_r (x)=\frac{|x|K}{\pi}
\label{nb}
\enq
In terms of the quantities defined,  the location of the zeros in the
 $t$-plane can be written as
\beq
t'_n= t_r  \frac{n_r(x)}{n}, ~~t"_n=t_d (1-\frac{n_r (x)}{n})
\label{t_n 1}
\enq
Note that we shall always have zeros on the upper half of the complex
 $t$-plane, since $n$ can be a negative integer, but zeros on the lower
half of the complex plane can exist only for $n$ a positive integer
 which satisfies
\beq
 n_r (x) > n
\label{nrn}
\enq
This implies that when
\beq
n_r(x)~<~1
\label {nr1}
\enq
there are no zeros on the lower half of the complex plane (since \er{nrn}
cannot be satisfied for any positive integer). Therefore,
the reciprocal relations in \er {RRim} and \er{RRre} can be applied to
obtain the additional reflection-induced phase through the change in
 amplitude of the wave. Thus under this condition the reflection-induced
phase is an observable quantity.

The above inequality amounts to the following:
\beq
|x|p~<~\pi\hbar
\label  {xp}
\enq
The  above two inequalities are among the central results of this work and are
termed "Analyticity thresholds for  a reflected particle".
They can be achieved if the distance of observation $|x|$ from the position of
 the barrier is short, or if the mean momentum $p$ of the particle is small.
The latter condition can be achieved, as we shall see, for a particle with a microscopic
mass or for an extremely slowly moving projectile. The momentum-position
uncertainty relations are not violated by \er {xp}, since $p$ is not the measured
momentum of the particle, only a parameter in the preparation of the projectile. Likewise,
$x$ is a parameter of the measurement, whose outcome is a spread-out function
(the wp modulus).

In terms of the parameters introduced in this section, the difference function
can be written more simply as
\beq
\chi'(t)~=~ 1-\exp \big[i\pi n_r(x)\frac{t_r - i t_d}{t-it_d}\big]
\label{chinew}
\enq

\section {Applications}
We consider three cases for which the wavepacket in \er{Psi} can serve
as prescriptions.
\subsection {An electron}
The wp of this can be characterized by the following parameters (all in atomic
units):
$m=1 (=m_e $, electronic mass), $a=-5$, $v$ (velocity)$=K/m=2$, $\Delta=2$,
$x=-1.5$

 Then $|x||K|=3$, $E=2$, $t_r=\frac{m|a|}{K} =2.5$, $\pi n_r(x)=|x||K|=3$,
$t_d=8$,
\beq t'_n=\frac{7.5}{\pi n}, ~~~t"_n =8(1-\frac{3}{\pi n}) \label
{electront} \enq
 where n is zero or a positive or negative integer. It then follows that
  $t"_n<0$ for all $n$. Thus, all zeros of the difference function
$\chi(t)$ lie in the upper half of the
 $t$-plane. By consequence  $\ln \chi(t)$ is analytic in the lower half-plane
 and vanishes on a large semi-circle there. This ensures the
 validity of the reciprocal relations shown in \er {RRim} and \er {RRre}.

\begin{figure}
\vspace{5.6cm}
\begin{picture}(1,1)
\end{picture}
\includegraphics{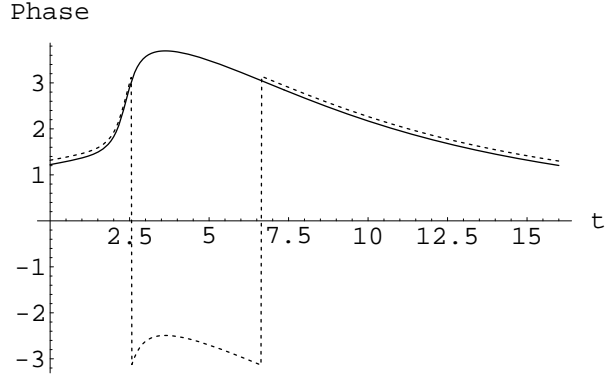}
\caption {The "analytical phase". This is the part of the phase coming
from the analytic difference function $\chi (x,t)$ (here plotted for an
electron against time $t$ in the vicinity of the reflection time $t_r$
and for $x=-1.5$, all in atomic units). The phase shown by full lines is
calculated from the log-modulus, using one of the two conjugate reciprocal
relations. The curve drawn by broken lines is calculated directly as the
argument of the complex function of $\chi (x,t)$. It shows jumps of $\pm
2 \pi$. } \label{kkphase}
\end{figure}

 We illustrate the use of the reciprocal relations in Figure \ref{kkphase}.
 The curve  shown by full lines is obtained by calculating the
argument indirectly, from the modulus through \er {RRim}. The broken line curve,
shown for comparison and displaced by a tiny amount for clarity, is computed
 directly from the expression
 in \er {chi}. It is clear that the two curves represent the same quantity
calculated in two different ways.

Apart from verifying the analytical properties of $\chi(x,t)$, the
agreement between the curves in Figure 2 provides a further instance for
the possibility, achievable under suitable circumstances, that for a wp
the analytic part of the phase is precisely given by the values of the
wave-function modulus for real values of the time. (The "analytic" part
is obtained from the total, physical phase by subtracting from the latter
those quantities that do not vanish at $t=\pm\infty$.)
 This part of the phase is "observable", indirectly, through the modulus.

The value of $x$ chosen for this case, namely .08 nanometers, is near
the analyticity threshold value for the chosen mean particle momentum.

\subsection {A molecule}
As the next example (and still staying inside the microscopic domain) we
take an impacting water molecule (molecular weight: 18)  with the following
 kinetic parameters  expressed in atomic
units: $m=3.6~ 10^4$, $a=-10$, $v$ (velocity)$=K/m=2.8~ 10^{-4}$, $K=10$,
$x=-4$. For the wave
packet width we use a value taken from ref. 18, namely  $\Delta=.3$
 (also in atomic units). This is of the order of the zero point motion amplitude
of the nuclei. The coordinate $x$ represents the position of the center of mass
of the molecule, with other internal degrees of freedom considered fixed during
the motion.

\begin{figure}
\vspace{6cm}
\begin{picture}(1,1)
\end{picture}
\includegraphics{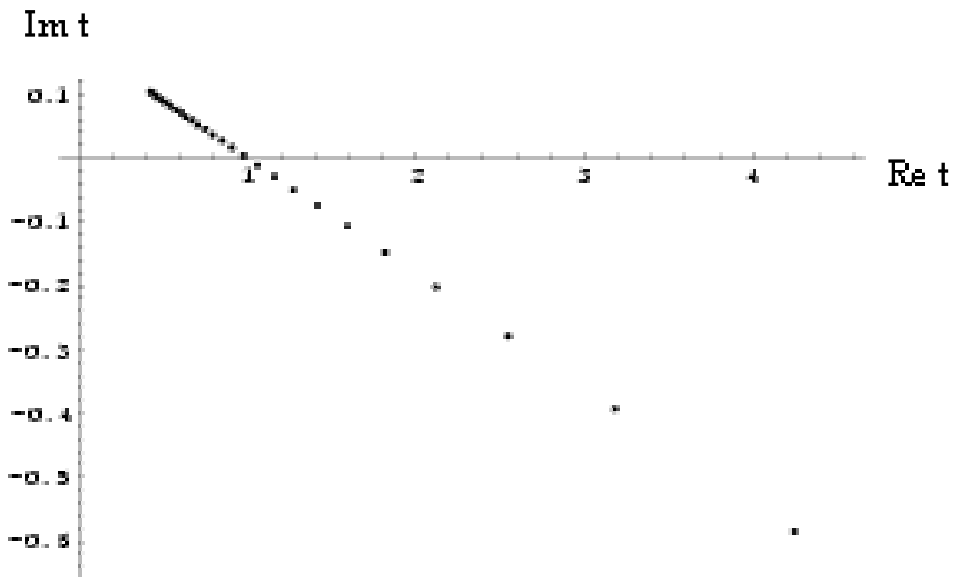} \caption {Argand plot in the complex $t (=t'+it")$ plane for
the location of  zeros in the difference function for a reflected water
molecule. The physical parameters are as given in the text for a
molecule. For these there are an infinite number of zeros in the
upper-half and 12 zeros in the lower-half of the complex t-plane. We show
about 30 zeros lying nearest to the real axis. The zeros are plotted in
units of the reflection time $t_r$, defined in the text and (with the
present choice of parameters) having a value of $3.6~10^4$ in atomic
units. } \label{molzeroes}
\end{figure}

 We further take $|x||K|=40$, $E=1.4~ 10^{-3}$, $t_r=\frac{m|a|}{K} =3.6~10^4$,
 $\pi n_r(x)=|x||K|=40$, $t_d = 3.1~10^3$,
\beq t'_n=\frac{1.4~ 10^6}{\pi n}, ~~~t"_n = 3.1~10^3 (1-\frac{40}{\pi
n}) \label {watert} \enq The zeros of the difference function are shown
in Fig. \ref{molzeroes}.

for values of $n$ in the neighborhood of the sign change in the imaginary
part $t"_n$, namely near $n=13 \approx 40/\pi$.

The position- threshold for analyticity in the case of a molecule that has the
 quoted dynamic  parameters is .02 nanometer, which is hardly accessible to measurements.
Thus for a molecule the determination of phase from modulus values requires the
consideration of the Blaschke term, as noted in $~{16}$.

\subsection {A classical object}
We express the parameters of the wp, now in {\it mgs units}, as follows: $ m=1 $,
$v=100 $, $mv=\hbar K=10^2$, ($\hbar \approx 10^{-31}$), $K= 10^{33}$, $|x|=0.1$,
$E=.5~10^2$ and in terms of an angular frequency, (in inverse second units) =
$E/\hbar =.5~10^{33}$, $t_r=10^{-2}$,  $\pi n_r (x)=|x||K|=10^{32}$, $ t_d= 2~10^9$.

The choice of the initial wp width $\Delta $ requires some thought, since it is
not a usual quantity for a classical object. A lower limit is clearly the
 zero point motion amplitude for a {\it single} vibrational mode in the object.
This is similar to that used for a molecule, above, and amounts to $\Delta= 10^{-11}$,
in the now used mgs units. It could be argued that the large number of vibrational
modes in the solid (of the order of $10^8$ modes in the direction of motion), would
demand  a larger number, e.g. through multiplication of the previous conjecture
 by the square root of the number of modes, leading to $\Delta= 10^{-7}$. We
shall see, however, that this latter number does not change qualitatively
the essential conclusions reached in this section, namely, that at about
the instant of the reflection, the wp acquires a very large phase shift, due
 to  the difference function.

Thus, with the former, smaller choice for $\Delta$,
 \beq t'_n=\frac{10^{30}}{\pi n}, ~~~t"_n = 2~ 10^{9}
(1-\frac{10^{32}}{\pi n}) \label {classt} \enq
 Because of the very large numbers involved, the discussion that follows
 will be in terms of orders of magnitudes.

\subsubsection {The reflectional phase shift}
We first rewrite the difference function in \er{chinew} as \beq
\chi'(t)~=~ 1- e^{ \big[ i\pi n_r(x)\frac{1 +\frac{it_r}{t_d}}{1
+\frac{it} {t_d}}\big]} \label{chiclas1} \enq so that the time-quotients
in the exponents are small and whose powers beyond the first can be
ignored. We first compute the phase-change (arising from the difference
function only) acquired during the full motion. Initially at $t=0$ we
compute (using the linear approximation in small quantities) \beq
\chi'(0)~=~ 1- e^{-i\pi n_r(x)} e^{-\pi \frac{n_r(x)t_r}{t_d}}= 1-
e^{-i\pi n_r(x)} \exp \big[-\frac{1}{2}10^{32-2-9}\big] \approx 1
\label{chiinit} \enq Thus, the phase is zero initially. Long after
 the reflection, but before disintegration of the wp $(t_d > t >>t_r)$
\ber
 \chi'(0)&=& 1- e^{i\pi n_r(x)} e^{\pi
\frac{n_r(x)t}{t_d} }\approx 1- e^{i\pi n_r(x)}
\exp[10^{30}]\
\nonumber \\
 &\approx&  e^{i\pi n_r(x)}\exp[10^{30}]\
\label{chifin}
\enr
The phase acquired is thus approximately $\pi n_r(x)=|x|K \approx 10^{32}$. What is
essential to note is that this phase scales with the momentum of the classical
particle $K$.
\subsubsection {The phase at reflection}
Let us next consider the phase change at about the reflection time $t_r$ (measured at
a point positioned at $|x|= 0.1 $ meter before the barrier). To this end we expand the exponent
 for $t\approx t_r$ and obtain
\beq
\chi'(0)~=~ 1- e^{i\pi n_r(x)} e^{\pi n_r(x) \frac{t-t_r}{t_d}} \approx
1-e^{i\pi n_r(x)} \exp[\frac{1}{2}~10^{21}(t-t_r)]
\label{chiref}
\enq
It is clear that the amplitude of the second (complex) term changes around
$t =t_r$ from a very small number to a very large one. The rate of change is
easily  calculated to be
\beq
\frac {|x|K\hbar}{\frac{1}{2}m\Del^2} = \frac{|x|v (the~velocity)}
{\frac{1}{2}\Del^2}= 2~10^{23} (radians~ per~ second)
\label {rate}
\enq
It should be noted that the astronomical large value of this rate will still remain
large $(\approx 10^{15})$ when the initial value of the wp width is increased
 $10^4 $-fold.

On the other hand, the rate of phase-change of a freely evolving
classical wave packet (in the absence of barrier) will  be larger
$ (\approx 10^{44})$ by many orders of $10$. This change remains  also for the
reflected particle, having regard to the factorization of $\Psi$ as described
 at the beginning of section 3. The pigmied reflection-induced phase may thus
be thought to be devoid of any physical significance. However, the former phase
is the {\it total} phase of the wp and is not in general observable, whereas
the reflection-induced phase is the relative phase between components of the
wp, which can be measured by interferometric methods.

While it may be said that the extremely large numbers met with in
this section for a macroscopic particle give the results a surrealistic look,
one notes that similar extremely high rates ($10^{19} s^{-1}$) are at the base
 of some proposals for wave function collapse as being due to
phase-decoherence $~^{19,20}$. We further discuss
 this connection in the concluding section.

The analyticity threshold  for the position of the observation has, for a macroscopic
projectile, an extremely low, sub-nucleonic value and is not realistic.

\section{Conclusion}
We have presented an analytic formulation for  an elementary one-
dimensional scattering process of a microscopic wave-packet, perhaps the
most elementary one for which an exact, analytic solution is available.
It has been shown that under circumstances that are realizable for an
electron, reciprocal relations hold between the phase and the modulus of
the scattered particle's wave function. This supplements our previous
demonstrations of the existence of reciprocal relations for localized,
bound states $~^{8-10}$. In those publications
 we have noted some other relations that connect the phase with modulus,
 in particular, the equation of continuity. However, this equation is a
partial differential equation and does not uniquely reconstruct the phase,
 even when the modulus is completely given. Thus, even for a problem in
only one spatial dimension, the addition to the position-derivative of the
 phase of a quantity $\phi
(t)/|\psi(x,t)|^2$, where $\phi(t)$ is  arbitrary, will satisfy the continuity
equation, while in higher dimensions a much larger family of functions will
do so. In contrast, the reciprocal relations give the analytic part uniquely
 from the modulus.

The physical basis for these relations was elucidated in $~^{14,15}$
, as being due to the lower-boundedness of
the energies. When the analyticity requirements are not fully satisfied, e.g.,
through  there being nodal points in the wave function, the phase is still
obtainable from the modulus, the latter being given as function of
the {\it complex} time. When the analytic properties postulated in this
article hold fully, it is sufficient to know the modulus as function of real
time, i.e., along the real t-axis.

 In this paper we have obtained
threshold relations which delimit the straightforward application
of the reciprocal relations. It will be of interest
to extend the theory to three dimensional scattering problems, to
finite-height barriers and to other cases.

For a (rigid) molecule impinging on an infinite barrier, we have found that
one should expect zeros of the "difference function" in the lower-half of
 the  complex time-plane, such that additional (so called, Blaschke) terms are
required to correlate the modulus with the phase changes.

When the barrier is not infinite, the algebraic, image wave-function solution
 used in this work is not applicable. For that situation we have developed a
method that employs a transfer matrix for each momentum state and we are in the
process of obtaining results from this method.

 In a further application of the formalism,  the reflection  of a classical
 particle from an infinite barrier is characterized by an extremely rapid
 rate of growth of the wave-function phase and by its attaining a very high
 value. Having taken  for the particle's mass  1g and for its velocity
100 m/s, we obtain  for the rate of  phase change at reflection, values of
the order of $10^{23}$ radians per second. The precise values, coming from the
difference function in \er{chi}, depend on the distance from the barrier and
 from the starting point of the particle.

The acquisition by classical systems of very large phases
  within a very short time period is likely to be quite general.
 It is expected that it is rooted in the reciprocal relations between the
 two quantities (moduli and phases) or, equivalently,  in the circumstance
 that they are  parts of the same analytic function. Thus as the
 $logarithm$ of the modulus increases (numerically), then so does the
 phase.

 While it is of a speculative nature, one would imagine a similar
 phenomenon to occur during a wave-function
 collapse.   It is widely held that the collapse is caused by a (phase-)
 decoherence process, which takes place while the quantum component is
coupled to (entangled with) the classical component (e.g. ref. 11).
 On the basis of our results  that the sudden switching
from one (positive mean momentum) state to another (negative mean momentum)
state by a macroscopic object causes the sudden acquisition of a large phase,
 we might expect a similar phase-increase to occur in the macroscopic part of
the combined quantum-classical system. A tiny variation in
this macro-phase will suffice to cause decoherence in the total wave
-function. However, a
detailed description of this process requires further work, probably through
use of one the proposed models for decoherence $~^{11}$.

\section {References}
~

\noindent
 (1) Messiah, A. {\it Quantum Mechanics} Vol II {North Holland, Amsterdam
1961) Chapter XVII. \\

\noindent
(2) R. Winter, R. {\it Phys. Rev.} {\bf 1961}, {\it123},  1503. \\

\noindent
(3) Dicus,D.A.;  Repko, W.A.;  Schwitters, R.F.; T.M. Tinsley, T.M.
{\it Phys. Rev.} {\bf A 202}, {\it 65}, 032116.\\

\noindent
(4) Kalbermann, G. {\it J. Phys. A: Math. Gen.} {\bf 2001}, {\it 34}, 36465;
{\it quant-phys}/02030036 (5 July  2002).\\

\noindent
(5) Savalli, V.;  Stevens, D.;  Esteve, J.; Featonby, P.D.; Josse,
V.;  Westbrook, N.; Westbrook C.I.;
 Aspect, A. {\it Phys. Rev. Lett.} {\bf 2002} {\it 88}, 250404.\\

\noindent
(6)  Berry, M.V.; {\it J. Phys. A: Math. Gen.} {\bf 2002}, {\it 35}, 3025.\\

\noindent
(7)  Bies, W.E.;  Heller, E.J.; {\it J. Phys. A: Math. Gen.}
 {\bf 2002}, {\it35}, 5673.\\

\noindent
(8) Englman, R.;  Yahalom, A.; Baer, M. {\it Phys. Lett. A} {\bf 1999},
{\it 251}, 223.\\

\noindent
(9)  Englman, R.;Yahalom, A.; {\it Phys. Rev. A } {\bf 1999}, {\it 60}, 1802.\\

\noindent
(10) Englman, R.; Yahalom, A. {\t Adv. Chem. Phys.} {\bf 2002}, {\it 124}, 197.\\

\noindent
(11)  Omnes, R.; {\it Phys. Rev. A}  {\bf 2002}, {\it 65}, 052119.\\

\noindent
(12) Adler,  S.; {\it J. Phys. A: Math. Gen.} {\bf 2002}, {\it 35}, 841.\\

\noindent
(13) Tomonaga, S.-I. {\it  Quantum Mechanics} Vol. II (North Holland,
Amsterdam, 1966} Sections 41 and 61.\\

\noindent
(14) Khalfin, L.A. {\it Soviet Phys. JETP}  {\bf 1958}, {\it 8}, 1053.\\

\noindent
(15) Perel'man, M.E.; Englman, R. {\it Modern Phys. Lett.} {\bf 2001}, {\it14}, 907.\\

\noindent
(16) Toll, J.S.; {\it  Phys. Rev.} {\bf 1956}, {\it 104}, 1760.\\

\noindent
(17) Gogtas, F.; Balint-Kurti, G.G.; Offer, A.R. {\it J. Chem. Phys.}
{\bf 1996}, {\it 104}, 7927.\\

\noindent
(18) Ghirardi, G.; Rimini, A.; Weber, T. {\it Phys. Rev. D}, {\bf 1986}, {\it34}, 470.\\

\noindent
(19) Pearle, P.;  Squires, E. {\it Phys. Rev. Lett.} {\bf 1994}, {\it 73} 1.\\

\end{document}